\pdfoutput=1
\RequirePackage{ifpdf}
\ifpdf % We are running pdfTeX in pdf mode
\documentclass[pdftex]{sigma}
\else
\documentclass{sigma}
\fi

\usepackage{epic,accents}

\def \h#1{\widehat{#1}}
\def \t#1{\widetilde{#1}}
\def \b#1{\overline{#1}}

\def \th#1{\widehat{\widetilde{#1}}}
\def \hb#1{{\widehat{\overline{#1}}}}
\def \bh#1{{\widehat{\overline{#1}}}}

\def \tb#1{\widetilde{\overline{#1}}}
\def \bt#1{\widetilde{\overline{#1}}}
\def \dh#1{\underaccent{\hat}{#1}}

\def \dt#1{\underaccent{\tilde}{#1}}

\numberwithin{equation}{section}

\begin{document}

\allowdisplaybreaks

\renewcommand{\thefootnote}{$\star$}

\renewcommand{\PaperNumber}{061}

\FirstPageHeading

\ShortArticleName{Soliton Taxonomy for a Modif\/ication of the Lattice Boussinesq Equation}

\ArticleName{Soliton Taxonomy for a Modif\/ication\\ of the Lattice Boussinesq Equation\footnote{This
paper is a contribution to the Proceedings of the Conference ``Symmetries and Integrability of Dif\/ference Equations (SIDE-9)'' (June 14--18, 2010, Varna, Bulgaria). The full collection is available at \href{http://www.emis.de/journals/SIGMA/SIDE-9.html}{http://www.emis.de/journals/SIGMA/SIDE-9.html}}}

\Author{Jarmo HIETARINTA~$^{\dag \ddag}$ and Da-jun ZHANG~$^\S$}
\AuthorNameForHeading{J.~Hietarinta and D.J.~Zhang}

\Address{$^{\dag}$~Department of Physics and Astronomy, University of
  Turku, FIN-20014 Turku, Finland}
\EmailD{\href{mailto:jarmo.hietarinta@utu.fi}{jarmo.hietarinta@utu.fi}}
\URLaddressD{\url{http://users.utu.fi/hietarin/}}

\Address{$^{\ddag}$~LPTHE / CNRS / UPMC, 4 place Jussieu 75252 Paris CEDEX   05, France}

\Address{$^\S$~Department of Mathematics, Shanghai University, Shanghai 200444, P.R. China}
\EmailD{\href{mailto:djzhang@staff.shu.edu.cn}{djzhang@staff.shu.edu.cn}}

\ArticleDates{Received May 24, 2011, in f\/inal form July 01, 2011;  Published online July 06, 2011}

\Abstract{Integrable multi-component lattice equations of the Boussinesq family
have been known for some time. Recently some new equations of this
type were found using the Consistency-Around-the-Cube approach. Here
we investigate one of these models, B-2, and in particular the
consequences of a nonzero deformation parameter $b_0>0$, which allows
special kinds of solitons in the parameter range $-b_0/3<k<b_0$.}


\Classification{35C09, 37K10, 39A14}

\renewcommand{\thefootnote}{\arabic{footnote}}
\setcounter{footnote}{0}

\section{Introduction}
The lattice Boussinesq equation (lBSQ) reads \cite{TN04}
\begin{subequations}\label{DB}
\begin{gather}
\t y - x\t x+z=0,\qquad \h{\t y} - \h x\h{\t x}+\h z=0,\label{DB-a}\\
\h y - x\h{x}+z=0,\qquad \h{\t y} - \t x\h{\t x}+\t z=0,\label{DB-b}\\
y - x\th{x}+\th{z}-\frac{p-q}{\t{x}-\h{x}}=0,\label{DB-c}
\end{gather}
\end{subequations}
where we have used the standard shorthand notation, e.g., $\t
x=x_{n+1,m}$, $\h z=z_{n,m+1}$, and where~$p$ and~$q$ are parameters
associated with the~$n$ and $m$ directions, respectively.  Equations
(\ref{DB-a}), (\ref{DB-b}) are def\/ined on the {\em edges} of the
elementary square of the Cartesian lattice, while~\eqref{DB-c} is
def\/ined on the square itself. These equations can be naturally
extended to a third dimension and the extension is
Consistent-Around-the-Cube (CAC)~\cite{TN04}.
(For further information about CAC see, e.g.,  \cite{cac-BS,cac-NW,cac-N}.)

The deformation called B-2 in \cite{JH11} is given by
\begin{gather}\label{DB-cp}
y - b_0(\th x-x)-x\th{x}+\th{z}-\frac{p-q}{\t{x}-\h{x}}=0.\tag{$\ref{DB-c}'$}
\end{gather}
The set (\ref{DB-a}), (\ref{DB-b}), (\ref{DB-cp}) also has the CAC
property~\cite{JH11}. By changing the sign of $x$, $p$, $q$, if necessary, we may assume
without loss of generality that $b_0\ge0$. This will be used in the following.

In order to understand the physical meaning of the new parameter $b_0$
in~\eqref{DB-cp} we construct its soliton solutions. The method is
essentially the same as the one given in~\cite{HZ10}, where we
considered the standard lBSQ. We repeat here the salient parts of that
construction, concentrating on the dif\/ferences.

\section{Construction of the one-soliton solution}

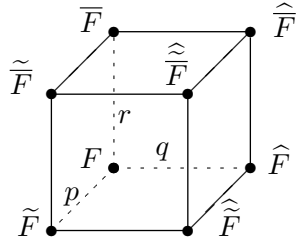
\begin{figure}[h]
\setlength{\unitlength}{0.0004in}
\hspace{2cm}
\hspace{4cm}
\begin{picture}(3482,3700)(0,-500)
%\put(-1200,1000){\makebox(0,0)[lb]{$(b)$}}
\put(450,1883){\circle*{150}}
\put(-100,1883){\makebox(0,0)[lb]{${\bt F}$}}
\put(1275,2708){\circle*{150}}
\put(825,2708){\makebox(0,0)[lb]{$\b F$}}
\put(3075,2708){\circle*{150}}
\put(3375,2633){\makebox(0,0)[lb]{${\bh F}$}}
\put(2250,83){\circle*{150}}
\put(2650,8){\makebox(0,0)[lb]{${\th F}$}}
\put(1275,908){\circle*{150}}
\put(1275,908){\circle*{150}}
\put(825,908){\makebox(0,0)[lb]{$F$}}
\put(625,408){\makebox(0,0)[lb]{$p$}}
\put(1825,1008){\makebox(0,0)[lb]{$q$}}
\put(1325,1508){\makebox(0,0)[lb]{$r$}}
\put(2250,1883){\circle*{150}}
%\put(2250,1883){\circle{220}}
%\put(2250,1883){\circle{80}}
\put(1950,2000){\makebox(0,0)[lb]{${\th{\b F}}$}}
\put(450,83){\circle*{150}}
\put(0,8){\makebox(0,0)[lb]{$\t F$}}
\put(3075,908){\circle*{150}}
\put(3300,833){\makebox(0,0)[lb]{$\h F$}}
\drawline(1275,2708)(3075,2708)
\drawline(1275,2708)(450,1883)
\drawline(450,1883)(450,83)
\drawline(3075,2708)(2250,1883)
\drawline(450,1883)(2250,1883)
\drawline(3075,2633)(3075,908)
\dashline{60.000}(1275,908)(450,83)
\dashline{60.000}(1275,908)(3075,908)
\drawline(2250,1883)(2250,83)
\drawline(450,83)(2250,83)
\drawline(3075,908)(2250,83)
\dashline{60.000}(1275,2633)(1275,908)
\end{picture}\vspace{-3mm}
\caption{The CAC-cube.\label{F:1}}
\end{figure}

In the construction we use the consistency cube of Fig.~\ref{F:1} in
an essential way. Here $F=(x,y,z)^T$ and $p$, $q$, $r$ stand for the
direction parameters of the three directions.
In practice consistency means that the equations on
the sides provide a  B\"acklund transformation between the bottom and
the top equations. In the present case the
side equations are
\begin{subequations}\label{B-yht}
  \begin{gather}
    \label{B-y}
    \b y  =  x\b x-z,\qquad \bt x = \frac{\t{z}-\b{z}}{\t{x}-\b{x}},
\qquad \hb{x} = \frac{\h{z}-\b{z}}{\h{x}-\b{x}}, \\
\tb{z}  = b_0(\bt x-x)+ x\tb{x}-y+\frac{p-r}{\t{x}-\b{x}},\\
\hb{z} = b_0(\bh x-x) + x\hb{x}-y+\frac{q-r}{\h{x}-\b{x}},
\label{B-z}
\end{gather}
\end{subequations}
along with some of their shifts.

\subsection{The background solution}

First one has to construct the background solution. This is
operationally the same as in \cite{HZ10}, but the parameter $b_0$ will
induce some changes.  We follow the f\/ixed-point idea
\cite{Sol-Q4-2007} which means that the variables on the bottom and
top square are the same.  Thus, omitting the bar-shift from~\eqref{B-yht}, the equations to solve are
\begin{gather*} y = x^2-z, \qquad
\t z-z=\t x(\t x-x),\qquad \h z-z=\h x(\h x-x),\nonumber\\
\t{z} = b_0(\t x-x)+x\t{x}-y+ \frac{p-r}{\t{x}-{x}} ,\qquad
\h{z} = b_0(\h x-x)+x\h x-y+  \frac{q-r}{\h{x}-x}.
\end{gather*}
The solution is given by
\begin{subequations}\label{0SS-HB}
\begin{gather}
x_0  = an+bm+c_1,\label{x0}\\
z_0  = \tfrac{1}{2}x^2_0+\tfrac{1}{2}\big(a^2n+b^2m+c_2\big)+c_3,\label{z0}\\
y_0  =\tfrac{1}{2}x^2_0-\tfrac{1}{2}\big(a^2n+b^2m+c_2\big)-c_3,\label{y0}
\end{gather}
\end{subequations}
where we have introduced new parameters $a$, $b$, which are related to $p$, $q$ by
\begin{gather}
a^3-b_0 a^2=p-r,\qquad b^3-b_0 b^2=q-r,
\label{abb0-pq}
\end{gather}
and $c_1$, $c_2$, $c_3$ are arbitrary constants (in~\cite{HZ10} a dif\/ferent
sign convention was used). Here the new parameter $b_0$ makes the
correspondence slightly more involved.

Since the parameters $a$, $b$ seem to be the natural ones for this
equation we may write~\eqref{DB-cp} using them obtaining
\begin{gather*}
y  -x\th{x}+\th{z}-\frac{a^3-b^3}{\t{x}-\h{x}}-b_0\left[(\th
    x-x)-\frac{a^2-b^2}{\t{x}-\h{x}}\right]=0.
%\label{HDB-c-ab-1}
\end{gather*}
Now we observe that the $b_0$ term is in fact related to lattice
potential KdV equation (lpKdV) given by
\begin{gather*}
(\th x-x)(\t{x}-\h{x})=a^2-b^2
\end{gather*}
In other words, \eqref{DB-cp} can be interpreted as a combination of
lBSQ~\eqref{DB-c} and~lpKdV.

\subsection{The one-soliton solution}

Next, in order to construct the one-soliton solution (1SS) we again use the
CAC-cube, now with variables $x$, $y$, $z$ at the bottom square
corresponding to the background solution and the bar-shifted variables
$\b x$, $\b y$, $\b z$ on the top square to the 1SS.  Thus, using
parametrization~\eqref{abb0-pq}, equations to solve are now
\begin{subequations}\label{1ss-yht}
  \begin{gather}
    \b y = x\b x-z,\qquad \bt x = \frac{\t{z}-\b{z}}{\t{x}-\b{x}},
\qquad \hb{x} = \frac{\b{z}-\h{z}}{\b{x}-\h{x}},\label{1ss-y} \\
\tb{z}  = b_0(\bt x-x)+ x\tb{x}-y+\frac{a^3-k^3-b_0(a^2-k^2)}{\t{x}-\b{x}},
\\
\hb{z} = b_0(\bh x-x) + x\hb{x}-y+\frac{b^3-k^3-b_0(b^2-k^2)}{\h{x}-\b{x}},
\end{gather}
\end{subequations}
where $k$ is the soliton parameter.

Following the method given in \cite{HZ10} we expand the 1SS as
\begin{gather}
(\b{x},\b{y},\b{z})=(\b{x}_0+X,\b{y}_0+Y,\b{z}_0+Z),
\label{bar-1SS}
\end{gather}
where $(\b{x}_0,\b{y}_0,\b{z}_0)$ is the bar-shifted $(x_0,y_0,z_0)$, i.e.,
\begin{subequations}\label{bar-0SS}
\begin{gather}
  \b x_0  = an+bm+k+c_1,\label{bx0}\\
  \b z_0  = \tfrac{1}{2}{\b x}^2_0+\tfrac{1}{2}\big(a^2n+b^2m+k^2+c_2\big)+
  c_3,\label{bz0}\\
  \b y_0  =\tfrac{1}{2}{\b x}^2_0-\tfrac{1}{2}\big(a^2n+b^2m+k^2+c_2\big)-c_3.\label{by0}
\end{gather}
\end{subequations}

With these def\/initions we f\/ind from \eqref{1ss-y} that $Y=x_0 X$. Thus
we only need to solve for~$X$,~$Z$, for which we have from~\eqref{1ss-yht}
\begin{alignat*}{3}
 & \t X  =  \frac{-\b{\t x}_0  X+Z}{X-(a-k)} ,\qquad &&  \t Z  =  \frac{-(\tb z_0+y_0+b_0
      x_0) X+(b_0+x_0) Z}{X-(a-k)},& \nonumber\\
&  \h X  =  \frac{-\b{\h x}_0  X+Z}{X-(b-k)}, \qquad &&
  \h Z = \frac{-(\bh z_0+y_0 +b_0      x_0)X+(b_0+x_0)Z}{X-(b-k)}. & %\label{BT-eqs}
\end{alignat*}
This system can be linearized by taking
$(X,Z)=(\frac{g}{f},\frac{h}{f})$, because then we can write it as
\begin{gather*}
\t\Psi=N\Psi,\qquad\h\Psi=M\Psi, \qquad %\label{Psi}
\Psi=(g,h,f)^T,
\end{gather*}
where
\begin{gather*}
N=\begin{pmatrix}
                     \tb x_0 &-1 &0\\
                     \tb z_0+y_0+b_0 x_0 &-b_0-x_0 &0\\
                     -1 & 0 & a-k
        \end{pmatrix},\qquad
M=\begin{pmatrix}
                     \hb  x_0 &-1 &0\\
                     \hb z_0+y_0 +b_0 x_0&-b_0-x_0 &0\\
                     -1 & 0 & b-k
        \end{pmatrix}.
\end{gather*}
These matrices satisfy the integrability condition $\h N M=\t M
N$. An important observation is that the $N$, $M$ matrices can both be
diagonalized using the matrix~$Q$ def\/ined by
\begin{gather*}
 % \label{dnm}
  Q(n,m)=\begin{pmatrix}
x_0(n,m)+b_0-\omega_2 & -1 & 0 \\
x_0(n,m)+b_0-\omega_1  & -1 & 0 \vspace{2mm}\\
\dfrac{-(x_0(n,m)+b_0- k)}{k(3k-2b_0)} & \dfrac{1}{k(3k-2b_0)} & 1
\end{pmatrix},
\end{gather*}
because then we have
\begin{gather*}
%  \label{dnm2}
  N=Q(n+1,m)^{-1} D(a) Q(n,m),\qquad  M=Q(n,m+1)^{-1} D(b) Q(n,m),
\end{gather*}
where
\begin{gather*}
D(a)=\begin{pmatrix}
a-\omega_2 & 0 & 0 \\
0 & a-\omega_1 & 0 \\
0 & 0 & a-k
\end{pmatrix}.
\end{gather*}
The entries $\omega_i(\neq k)$ appearing in the above equations are
the roots of
\begin{equation}\label{roots}
W^3-k^3-b_0\big(W^2-k^2\big)=0.
\end{equation}
This always has the root $\omega_0=k$, and if $b_0=0$ then
$\omega_i=k\zeta^i$, $i=1,2$, where $\zeta\neq 1$ is the cubic root of
unity.  The key observation is that changing $b_0$ changes the values
of the other roots~$\omega_i(k)$, and although the $\omega_i$, $i=1,2$
can still be complex conjugates of each other although not of
magnitude~$k$, we also have the novel possibility that all $\omega_i$
can be real. In fact
\begin{gather}\label{omega-j}
\omega_j(k)=\frac{1}{2}\Bigl[(b_0-k)+(-1)^j\sqrt{(b_0-k)(b_0+3k)}
 \Bigr],\qquad j=1,2.
\end{gather}
So $\omega_i$ are real when $-b_0/3<k<b_0$.
Using the above formulae we can construct $\Psi$:
\begin{gather*}
  \Psi(n,m)=Q(n,m)^{-1}D(a)^nD(b)^m Q(0,0)\Psi(0,0),
\end{gather*}
from which we f\/ind
\begin{subequations}\label{sol-1ss}
\begin{gather} \label{1ss}
f  =\sum_{\nu=0}^2 (a-\omega_\nu(k))^n(b-\omega_\nu(k))^m\rho^0_\nu,\\
g =\sum_{\nu=0}^2 (\omega_\nu(k)-k)
(a-\omega_\nu(k))^n(b-\omega_\nu(k))^m\rho^0_\nu,\\
h =x_0g+\sum_{\nu=0}^2 \big(\omega_\nu(k)^2-k^2\big) (a-\omega_\nu(k))^n
(b-\omega_\nu(k))^m\rho^0_\nu.
\end{gather}
\end{subequations}
Here, instead of $g_{00}$, $h_{00}$, $f_{00}$, we have introduced
new constants $\rho_\nu^0$, which are def\/ined by
\[
\rho_0^0=f_{00}-\frac{g_{00}(c_1+b_0-k)-h_{00}}{(\omega_1-k)(\omega_2-k)},\qquad
\rho_\nu^0=\frac{g_{00}(c_1+b_0-\omega_{\nu})-
  h_{00}}{(-1)^{\nu}(\omega_1-\omega_2) (\omega_\nu-k)},\qquad \nu=1,2.
\]
Using (\ref{bar-1SS}), (\ref{bar-0SS}) we
can recover the 1SS as
\begin{subequations}
  \label{1ss-fin}
\begin{gather}
 x^{\rm 1SS}=x_0+\frac{\omega_0\rho_0+\omega_1\rho_1+
    \omega_2\rho_2}{\rho_0+\rho_1+\rho_2},\label{1ss-x}\\
    z^{\rm 1SS}=z_0+x_0
  \frac{\omega_0\rho_0+\omega_1\rho_1+
    \omega_2\rho_2}{\rho_0+\rho_1+\rho_2} +\frac{\omega_0^2\rho_0
    +\omega_1^2\rho_1+ \omega_2^2
    \rho_2}{\rho_0+\rho_1+\rho_2},\\
    y^{\rm 1SS}=y_0+x_0
  \frac{\omega_0\rho_0+\omega_1\rho_1+
    \omega_2\rho_2}{\rho_0+\rho_1+\rho_2}.
\end{gather}
\end{subequations}
Here $x_0$, $y_0$, $z_0$ were def\/ined in \eqref{0SS-HB} and
\begin{equation}
\rho_\nu(n,m)=\left(a-\omega_\nu(k)\right)^n
\left(b-\omega_\nu(k)\right)^m\rho_\nu^0,\qquad \nu=0,1,2.
\end{equation}
The $\rho_\nu$ correspond to the plane-wave factors (PWF) in the
continuous case and we use the same name here.

\section{The dif\/ferent types of solitons}
\subsection{Generic properties}
The soliton \eqref{1ss-fin}, when expressed in variables $x$, $y$, $z$,
has a linearly or quadratically growing background part~\eqref{0SS-HB}. In order to show the soliton behavior we ignore this
part and only discuss the part
\begin{equation}\label{def-kink}
U:=k+\frac{g}f=\frac{\omega_0\rho_0+\omega_1\rho_1+
    \omega_2\rho_2}{\rho_0+\rho_1+\rho_2}.
\end{equation}

From \eqref{def-kink} we can immediately see that the soliton is a
kink or anti-kink and that in the asymptotic region it will level of\/f
to some $\omega_i$, depending on the term that dominates in the
particular asymptotic region of $n$, $m$.  In comparing the terms $i$ and
$j$ it is useful to determine the line where they are equal, it is
given by
\[
n \log\left|\frac{a-\omega_i}{a-\omega_j}\right|+
m \log\left|\frac{b-\omega_i}{b-\omega_j}\right|=0.
\]
Then the $i$ term will dominate in the growing $n$ region with respect
to this line, if $|a-\omega_i|>|a-\omega_j|$, etc. Furthermore one
has to verify that the $i=j$ term is dominating over the $k$ term,
and therefore only a half-line is relevant.  In the generic case one
needs to draw three such half-lines, which results in three dif\/ferent
asymptotic regions.  Examples of these will be given below.

\subsection[The behavior of solitons for $b_0=0$]{The behavior of solitons for $\boldsymbol{b_0=0}$}

Let us f\/irst consider the classical case for which $b_0=0$. Then
$\omega_{\nu}=k\zeta^\nu$ where $\zeta=-\tfrac12(1+i\sqrt{3})$. For
simplicity let us assume that $0<k<a,b$. We have
\[
|a-\omega_\nu| = \sqrt{a^2+ak+k^2} \qquad \text{for} \ \ \nu=1,2,
\]
from which it follows that  $|\rho_1|=|\rho_2|$. Furthermore let us def\/ine
\[
\varphi_a(k)=\arccos\left(\frac{a+k/2}{\sqrt{a^2+ak+k^2}}\right)
\]
so that $a-\omega_\nu = \sqrt{a^2+ak+k^2}e^{i\varphi_a(k)}$.  Since
$\omega_1^*=\omega_2$ we can have real solitons if we choose
$(\rho_2^0)^*=\rho_1^0=\alpha e^{i\beta}$; without loss of generality
we can also take $\rho^0_0=1$.

\begin{figure}[t]
\centering

\includegraphics[width=50mm]{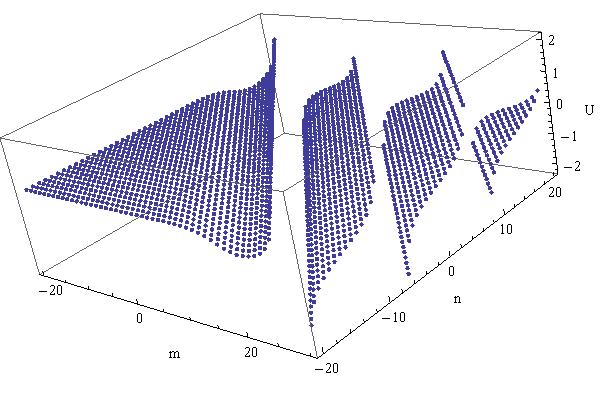} \qquad \quad
\includegraphics[width=60mm]{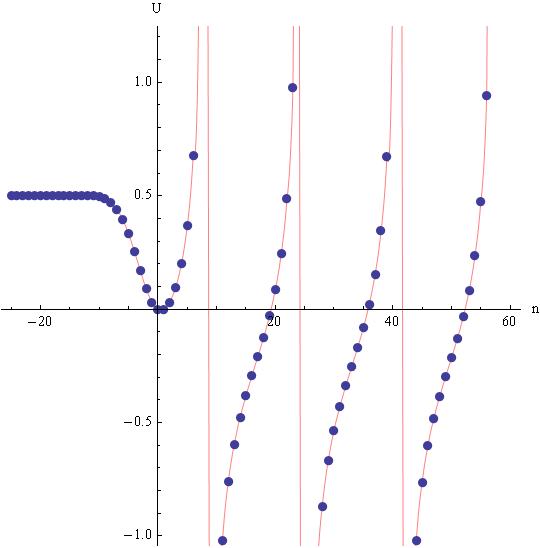}

\caption{Typical oscillation for the $b_0=0$ case. On the left a 3D
  view of $U$ of \eqref{def-kink}, on the right a~section at $m=0$. The
  parameters have values $b_0=0$, $a=b=2$, $k=0.5$, $\rho_\nu^0=1$.
  \label{Fosc}}
\end{figure}

We can now write the soliton part of \eqref{def-kink} in the
$b_0=0$ case as
\begin{equation}\label{1SSosc}
U=k \frac{A^nB^m+2\alpha\cos(\xi-\tfrac23\pi)}
{A^nB^m+2\alpha\cos \xi},
\end{equation}
where
\[
A=\frac{a-k}{\sqrt{a^2+ak+k^2}},\qquad B=\frac{b-k}{\sqrt{b^2+bk+k^2}},
\qquad\xi=n\varphi_a(k)+m\varphi_b(k)+\beta.
\]
Since $0<k<a,b$ we have $0<A,B<1$. Thus $A^nB^m\to +\infty$ as
$n,m\to-\infty$ and therefore in that direction $U\to k$. On the other
hand, $A^nB^m\to 0$ as $n,m\to+\infty$ and the behavior in that
direction is oscillatory since the cosine terms dominate. In the
continuous case there would be some values of the independent
variables for which the denominator actually vanishes. In the discrete
case we can only say that for large enough $n$, $m$ we can choose their
particular values so that~$\xi$ gets arbitrarily close to
$(N+1/2)\pi$. And then due to the extra $-2\pi/3$ in the numerator,
$U$ can take arbitrarily large (positive or negative) values.  Thus in
the $b_0=0$ case all single-soliton solutions are ef\/fectively singular
asymptotically. This is illustrated in Fig.~\ref{Fosc}.

\subsection[The behavior of solitons when $k<-b_0/3$ or $b_0<k$ ($b_0>0$)]{The behavior of solitons when $\boldsymbol{k<-b_0/3}$ or $\boldsymbol{b_0<k}$ ($\boldsymbol{b_0>0}$)}

  The solitons behave essentially the same way as for
$b_0=0$ also in the $b_0\neq 0$ case, provided that $k>b_0$ or
$k<-b_0/3$, i.e., $(k-b_0)(b_0+3k)>0$. The roots $\omega_i$, $i=1,2$ are
still complex conjugates, with $|\omega_i|^2=k(k-b_0)$, and
$|\rho_1|=|\rho_2|$.  The detailed expressions for variables
$A$, $B$, $\varphi_a(k)$, used in \eqref{1SSosc}, will just have a dif\/ferent
form. Let us assume $b_0<k<a,b$, then we have
\begin{gather*}
\omega_0=k,\qquad\omega_1=\tfrac12\left[b_0-k-i\sqrt{(k-b_0)(b_0+3k)}\right],
\\
\omega_2=\tfrac12\left[b_0-k+i\sqrt{(k-b_0)(b_0+3k)}\right],
\\
|a-\omega_i|=\sqrt{a^2+ak+k^2-b_0(a+k)} \qquad \text{for} \ \ i=1,2,
\\
\varphi_a(k)=\arccos\left(\frac{a+k/2-b_0/2}
{\sqrt{a^2+ak+k^2-b_0(a+k)}}\right).
\end{gather*}
For the plane-wave factors we have rations
\[
A=\frac{a-k}{|a-\omega_1|},\qquad B=\frac{b-k}{|b-\omega_1|},
\qquad A^2=1+\frac{a(b_0-k)+k(b_0-2a)}{a^2+ak+k^2-b_0(a+k)}<1,
\]
or more generally
\[
\left|\frac{a-\omega_\nu(k)}{a- \omega_0(k)}\right|^2
=\frac{a^3-k^3-b_0(a^2-k^2)}{(a-k)^3}\quad (>0).
\]

\begin{figure}[t]
\centering
\includegraphics[width=50mm]{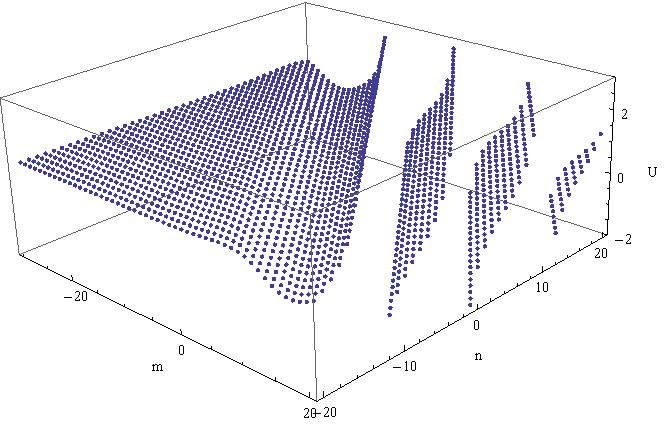} \qquad \quad
 \includegraphics[width=60mm]{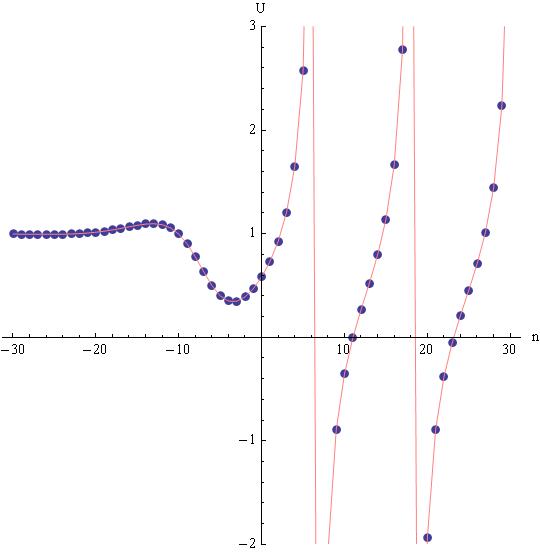}

\caption{Oscillatory behavior of $U$ for $b_0=1$, $a=b=4$, 
$\rho_\nu^0=1$, $k=1.75$, $m=0$.\label{F-osc}}
\end{figure}

One example of such a soliton is given in Fig.~\ref{F-osc}.

\subsection[The behavior of solitons when $-b_0/3<k<b_0$]{The behavior of solitons when $\boldsymbol{-b_0/3<k<b_0}$}

If $-b_0/3<k<b_0$ the situation changes completely, because the roots
$\omega_\nu(k)$ are now all real, and dif\/ferent in magnitude
\begin{gather*}
\omega_0=k,\qquad\omega_1=\tfrac12\left[b_0-k-\sqrt{(b_0-k)(b_0+3k)}\right],\\
\omega_2=\tfrac12\left[b_0-k+\sqrt{(b_0-k)(b_0+3k)}\right].
\end{gather*}
Now  dif\/ferent terms of $U$ can dominate
depending on the value of $k$.

\begin{figure}[t]
\centering \includegraphics[width=45mm]{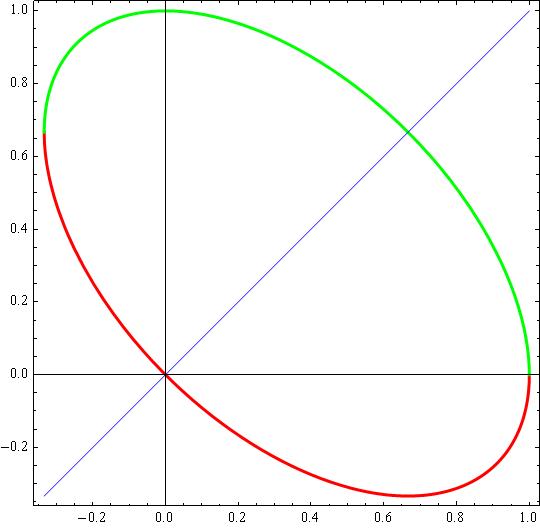}

\caption{The roots of \eqref{roots} as functions of $k$ for $b_0=1$:
  $w_2$ in green, $w_1$ in red, and $\omega_0=k$ in blue.}\label{Fw1w2}
\end{figure}

For the elements appearing in the PWF's We f\/ind
\begin{subequations}
\begin{alignat}{4}
& -\tfrac13b_0<k<0:\qquad && k<w_1<w_2,\qquad &&  a-k>a-w_1>a-w_2,& \label{tap1}\\
& 0<k<\tfrac23b_0:\qquad && w_1<k<w_2,\qquad && a-w_1>a-k>a-w_2,& \label{tap2}\\
& \tfrac23b_0<k<b_0:\qquad && w_1<w_2<k,\qquad && a-w_1>a-w_2>a-k. & \label{tap3}
\end{alignat}
\end{subequations}
The relationships of the middle column above can be seen if we plot
$k$, $w_1$, $w_2$ together, see Fig.~\ref{Fw1w2}.

There are now a variety of soliton types depending on the values of
soliton parameters $k$, $\omega_i$ with respect to the equation
parameters $a$, $b$. All the PWF's are positive, if $a,b>\max(k,w_2)$,
and if $a,b<\min(k,w_1)$ the negative sign can be canceled
from all terms. In these cases there will be no oscillations. If
however~$a$ and/or~$b$ falls between $\min(k,w_1)$ and $\max(k,w_2)$
then there will be oscillations and furthermore there can be a greater
variety in dominant asymptotic behavior.  We will next discuss some of
the possible cases.

\subsubsection[$a,b>\max(k,w_2)$ or  $a,b<\min(k,w_1)$]{$\boldsymbol{a,b>\max(k,w_2)}$ or  $\boldsymbol{a,b<\min(k,w_1)}$}

 If $a,b>\max(k,w_2)$ all PWF are positive. The term that
  dominates for large positive values of~$n$,~$m$ is $(a-k)^n(b-k)^m$ for~\eqref{tap1} and $(a-w_1)^n(b-w_1)^m$ in all other cases. For large
  negative values for both $n$, $m$ the dominating term is
  $(a-k)^n(b-k)^m$ for~\eqref{tap3} and $(a-w_2)^n(b-w_2)^m$ in all
  other cases. For other $n$, $m$ directions we must compare values more
  closely.  Furthermore, choosing a relatively large $\rho_i^0$
  will make the corresponding term dominate in some f\/inite region,
  although not asymptotically.

\begin{figure}[t]
\centering \includegraphics[width=60mm]{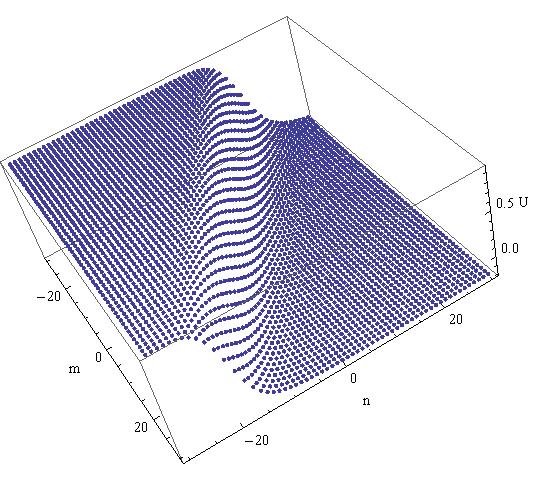} \qquad\quad
 \includegraphics[width=60mm]{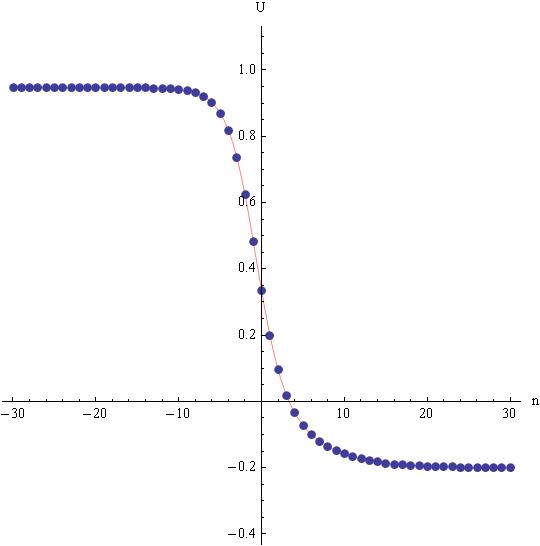}

\caption{When $a,b\ge \max(k,\omega_2)$ the soliton looks like
  a simple anti-kink. $b_0=1$, $k=-0.2$, $a=2$, $b=3$, $\rho_\nu^0=1$.}\label{Fig3a/b}
\end{figure}

  In the example of Fig.~\ref{Fig3a/b} the $a$, $b$ values are so large
  in comparison with $k$ that the solution looks like a simple
  anti-kink. Similarly for large negative values of $a$, $b$ we get a
  simple kink.

\begin{figure}[t]
\centering \includegraphics[width=60mm]{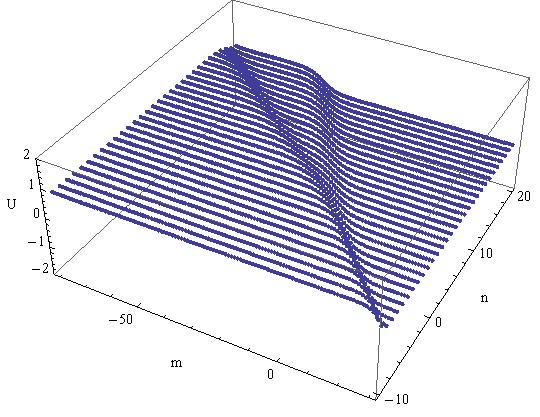} \qquad\quad
 \includegraphics[width=60mm]{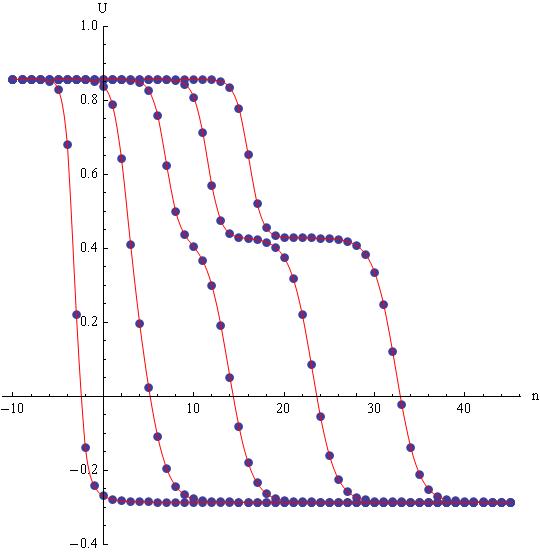}

\caption{A 3D view and sections for $m=-70,-50,-30,-10,10$ of an
  antikink with $b_0=1$, $k=\frac{3}{7}$, $a=1$, $b=2$, $\rho_\nu^0=1$.
All $a-\omega_i$, $b-\omega_i$ are positive.}\label{Fig5a/b}
\end{figure}

\begin{figure}[t]
\centering \includegraphics[width=60mm]{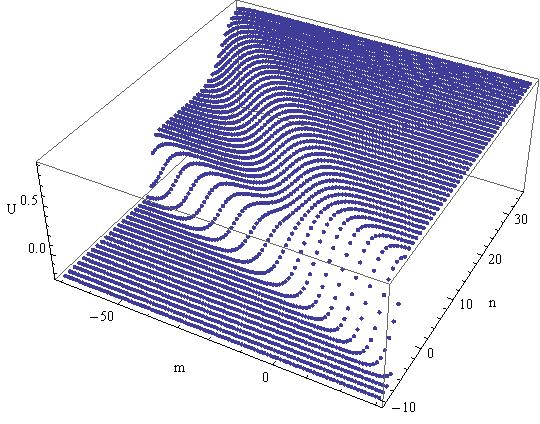}\qquad\quad
 \includegraphics[width=60mm]{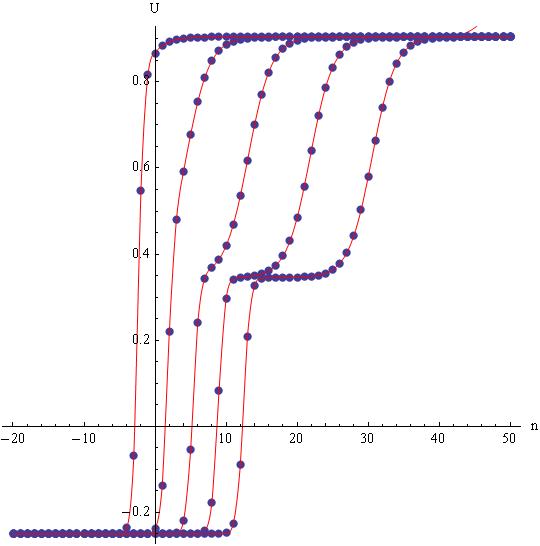}

\caption{A 3D view and sections for $m=-70,-50,-30,-10,10$ of a
  kink with $b_0=1$, $k=-0.25$, $a=-0.3$, $b=-1.5$, $\rho_\nu^0=1$. All $a-\omega_i$, $b-\omega_i$ are negative.
\label{Fig5e}}
\end{figure}

If $a$, $b$ are only moderately larger than $b_0$ the double soliton
nature of the kink starts to appear. This is illustrated in Fig.~\ref{Fig5a/b}. A kink is shown in Fig.~\ref{Fig5e}.

\begin{figure}[th!]
\centering \includegraphics[width=55mm]{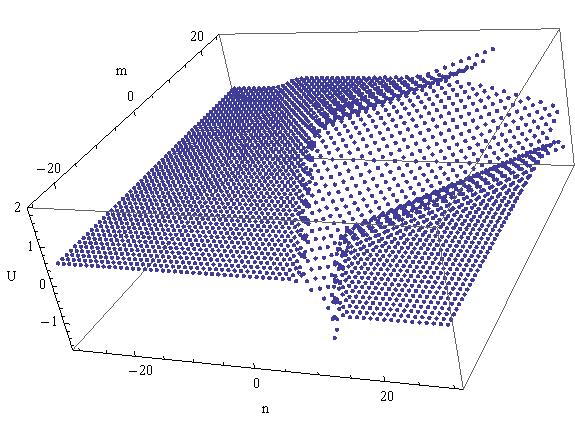}\qquad\quad
 \includegraphics[width=60mm]{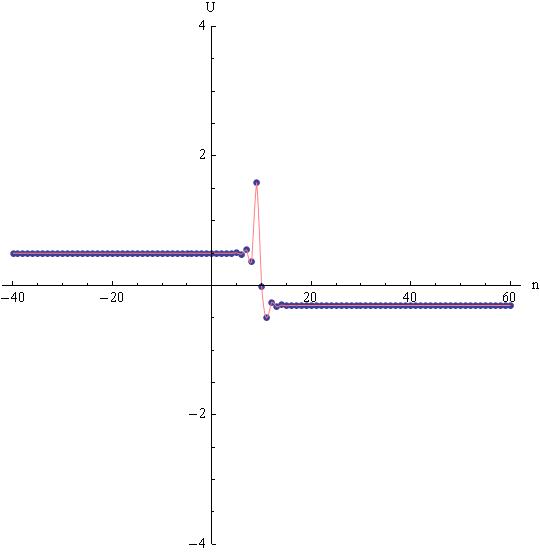}

 \includegraphics[width=60mm]{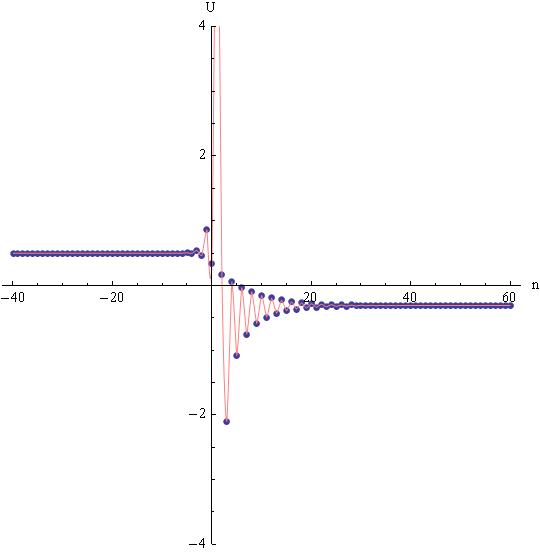} \qquad\quad
 \includegraphics[width=60mm]{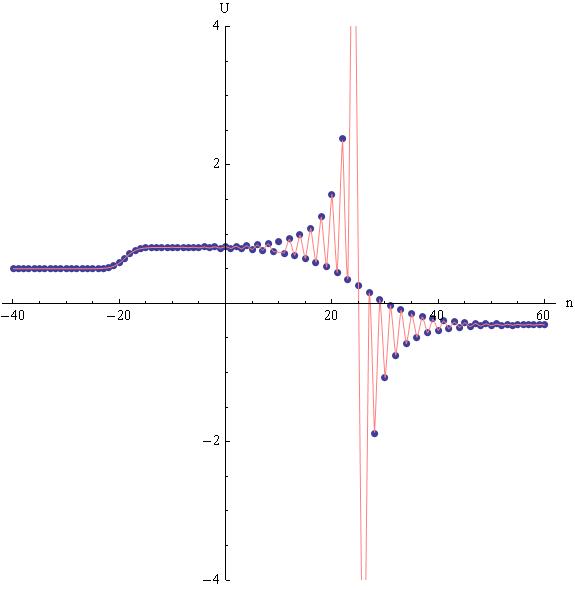}

\caption{Three-level 1SS with oscillations on the boundary of level
  $\omega_1$. 3D view with cross sections at $m=-20,0,25$.  The
  parameters are $b_0=1$, $k=0.5$, $a=0.2$, $b=0.3$, $\rho_\nu^0=1$.
  \label{Fig6}}
\end{figure}

\subsubsection[$\omega_1<a,b<\omega_0,\omega_2$]{$\boldsymbol{\omega_1<a,b<\omega_0,\omega_2}$}\label{342}

\begin{figure}[th!]
\centering\includegraphics[width=60mm]{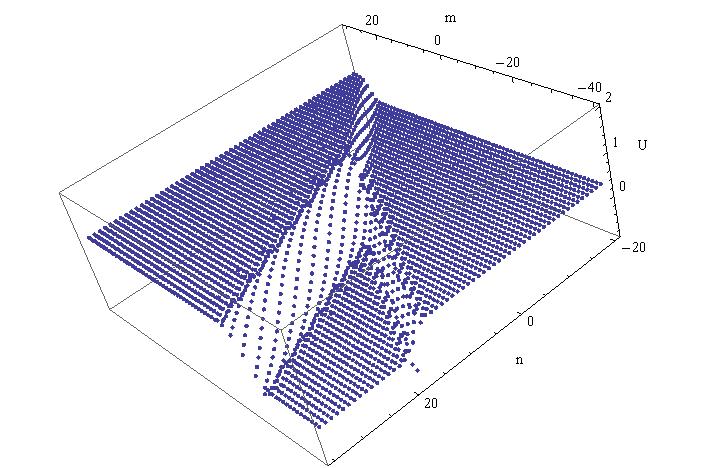}\qquad\quad
 \includegraphics[width=60mm]{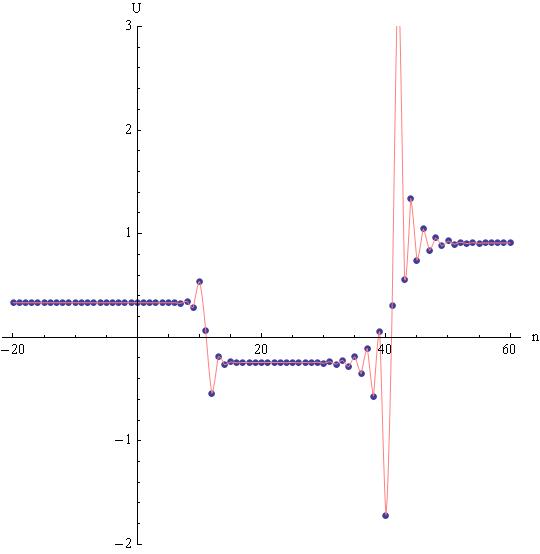}

\caption{3D view with cross section at $m=-25$. The parameters are
  $b_0=1$, $k=\frac{1}{3}$, $a=0.2$, $b=0.12$, $\rho_\nu^0=1$.}  \label{Fig8}
\end{figure}

For this choice the PWF's $\rho_0$ and $\rho_2$ have alternating signs
as functions of~$n$, $m$, but $\rho_1>0$. This means that at the line
where $|\rho_0|=\rho_1$ or $|\rho_2|=\rho_1$ there will be unbounded
oscillations in the asymptotic regions, while the behavior is smooth
around the line $|\rho_0|=|\rho_2|$. One such case is given in Fig.~\ref{Fig6}, for which the parameters are $b_0=1$, $k=0.5$,
$a=0.2$, $b=0.3$, $\rho_\nu^0=1$. It then follows that $\omega_0=0.5$,
$\omega_1=-0.309$, $\omega_2=0.809$ and $a-\omega_0=-0.3$,
$b-\omega_0=-0.2$, $a-\omega_1=0.509$, $b-\omega_1=0.609$,
$a-\omega_2=-0.609$, $b-\omega_2=-0.509$. The oscillatory
transition lines are at $n=-0.475  m$ for $|\rho_0|=|\rho_1|$ and $n=m$
for  $|\rho_2|=|\rho_1|$, while the transition line  $|\rho_0|=|\rho_2|$
at $n=-0.758  m$ is smooth.
Another example is illustrated in Fig.~\ref{Fig8}, the opening angle
to the lowest level is now smaller.

\subsubsection[$\omega_1<a<\omega_0<b<\omega_2$]{$\boldsymbol{\omega_1<a<\omega_0<b<\omega_2}$}

\begin{figure}[t]
\centering\includegraphics[width=60mm]{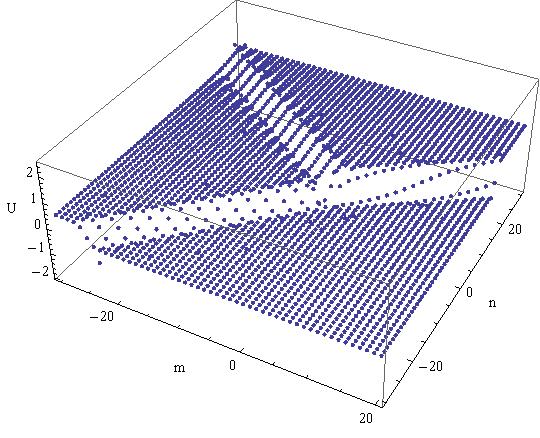}\qquad \quad
 \includegraphics[width=60mm]{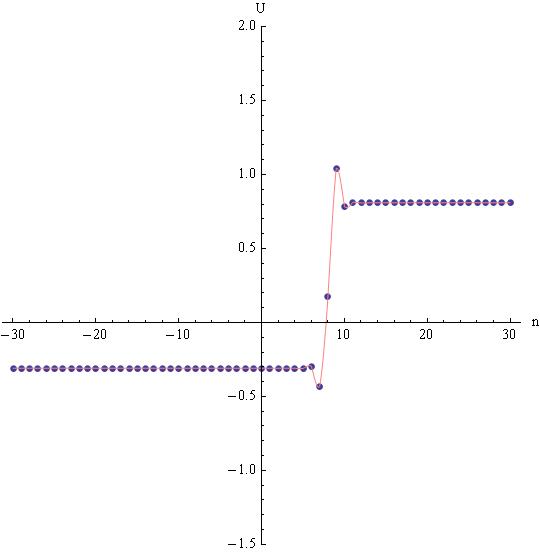}

 \includegraphics[width=60mm]{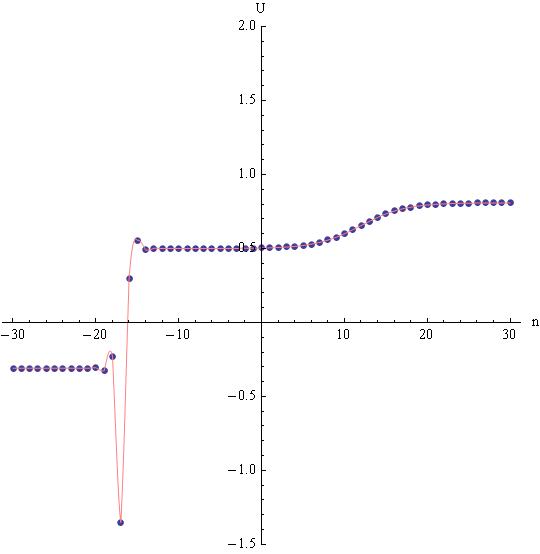}\qquad \quad
 \includegraphics[width=60mm]{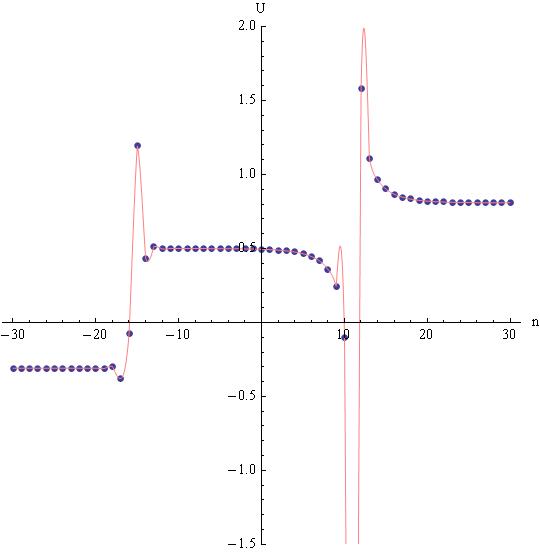}

\caption{Soliton solution with parameter values $b_0=1$, $k=0.5$,
$a=-0.18$, $b=0.63$, $\rho_\nu^0=1$. Above: 3D view with a section at
  $m=10$, below: two adjacent
  sections at $m=-14,-13$. The levels are, from left, 1, 0 and 2.
  The 0-2 transition line is smooth in $n$ for
  even $m$ and oscillatory in $n$ for odd $m$.}  \label{Fig7}
\end{figure}

The behavior is still more diverse when $a<\omega_0<b$. This is
illustrated in Fig.~\ref{Fig7}, where the parameters are $b_0=1$, $k=0.5$,
$a=-0.18$, $b=0.63$. The parameters $b_0,k,\omega_i$ are the same as
in Section~\ref{342} but since now $b>\omega_0$ the signs are dif\/ferent:
$a-\omega_0=-0.68$, $b-\omega_0=0.131$,
$a-\omega_1=0.129$, $b-\omega_1=0.939$,
$a-\omega_2=-0.989$, $b-\omega_2=-0.179$. Now all the transition lines
are oscillatory, the line $|\rho_0|= |\rho_1|$ oscillates as a function
of $n$, the lines $|\rho_0|= |\rho_2|$ and $|\rho_1|= |\rho_1|$ as
functions of $m$.

\subsubsection[$\omega_0=\omega_2$]{$\boldsymbol{\omega_0=\omega_2}$}

\begin{figure}[t]
\centering\includegraphics[width=55mm]{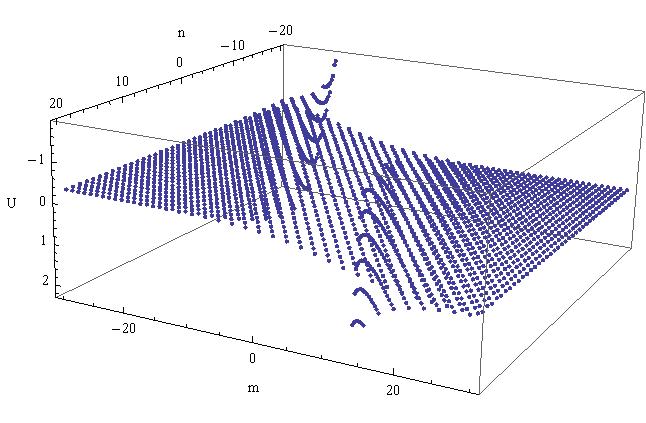}\qquad\quad
 \includegraphics[width=60mm]{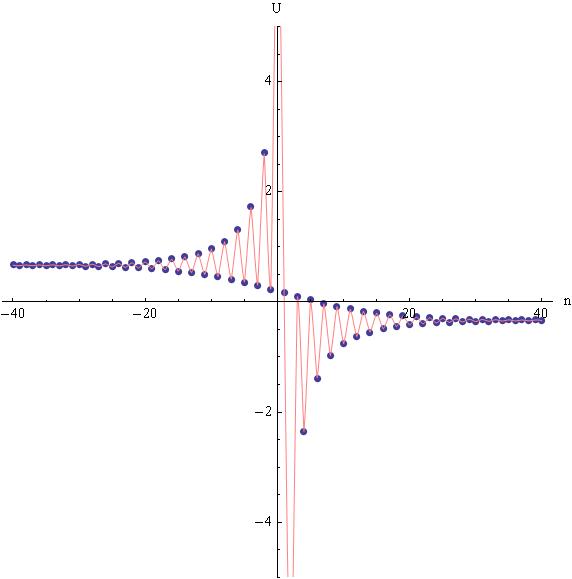}

 \includegraphics[width=60mm]{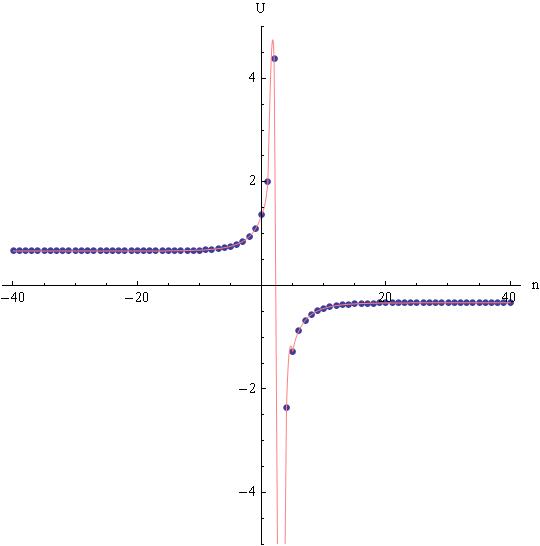}\qquad\quad
 \includegraphics[width=60mm]{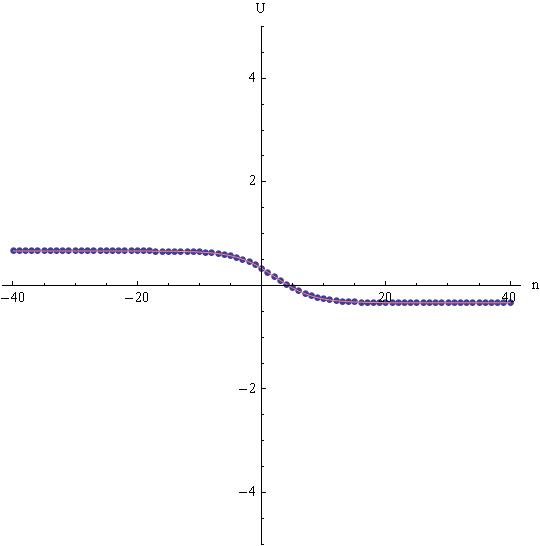}

\caption{A kink with oscillations. Here
  $b_0=1$, $k=2/3$, $a=0.2$, $b=0.12$ and therefore
  $\rho_0=\rho_2=(-1)^{n+m}|\rho_2|$, while $\rho_1>0$.
Top: 3D view and a section at $m=-3$, bottom: a section with
$n+m=1$ on the left and $n+m=0$ on the right.}  \label{Fig9}
\end{figure}

Finally we have an example for $\omega_0=\omega_2$. This leads to a
pure kink or antikink solution, because $\rho_0=\rho_2$.  This happens,
e.g., for $b_0=1,k=2/3$. Since $a-\omega_0<0$ but $a-\omega_1>0$ we
have oscillations on the transition region.  This is illustrated in
Fig.~\ref{Fig9}.

\section[Bilinearization, Casoratians and $N$-soliton solutions]{Bilinearization, Casoratians and $\boldsymbol{N}$-soliton solutions}

Above we have discussed only the 1SS. Although the f\/igures look
complicated one can obtain the multi-soliton solutions using a
procedure similar to \cite{HZ10}.  We only list the main results

By the dependent variable transformation
\begin{gather}
x=x_0-\frac{g}{f},\qquad z=z_0-x_0 \frac{g}{f}+\frac{h}{f},\qquad
y=y_0-x_0 \frac{g}{f}+\frac{s}{f},
\label{trans}
\end{gather}
we can bilinearize the B-2 lattice consisting of
$H\!\!B1$=\eqref{DB-a}, $H\!\!B2$=\eqref{DB-b},
$H\!\!B3$=\eqref{DB-cp} as
\begin{gather*}
  H\!\!B1 =\frac{\mathcal{B}_1}{f\t{f}},\qquad H\!\!B2  =\frac{\mathcal{B}_2}{f\h{f}},\\
  H\!\!B3
  =\frac{\mathcal{B}_3\mathcal{B}_4+(a-b)f\th{f}\mathcal{B}_4
+[a^2+ab+b^2-b_0(a+b)]\t{f}\h{f}\mathcal{B}_3}{(\t x-\h x)f\t{f}\h{f}\th{f}},
\end{gather*}
where  we have also used the parametrization \eqref{abb0-pq}, and
where the
bilinear equations are
\begin{subequations}
\begin{gather}
  \mathcal{B}_1 =\t{f}(h+ag)-\t{g}(g+af)+f\t{s}=0, \label{bil-HDB-a}\\
  \mathcal{B}_2 =\h{f}(h+bg)-\h{g}(g+bf)+f\h{s}=0,\label{bil-HDB-b}\\
  \mathcal{B}_3 =\t{f}\h{g}-\h{f}\t{g}+(a-b)(\t{f}\h{f}-f\th{f})=0,
\label{bil-HDB-c}\\
  \mathcal{B}_4 =\big[a^2+ab+b^2-b_0(a+b)\big](f\th{f}-\t{f}\h{f})\nonumber\\
  \phantom{\mathcal{B}_4 =}{} +
(a+b-b_0)(\th f g-f \th g)+\th{f}s+f\th{h}-g\th{g}=0.\label{bil-HDB-d}
\end{gather}
\label{bil-HDB-I}
\end{subequations}
The set of bilinear equations \eqref{bil-HDB-I} admits $N$-soliton
solutions in the following Casoratian form,
\begin{gather}
\label{casdef}
  f=|\h{N-1}|,\qquad\!  g=|\h{N-2},N|,
\qquad\!  h=|\h{N-2},N+1|,\qquad\! s=|\h{N-3},N-1,N|,\!\!\!
\end{gather}
composed of $\psi=(\psi_1,\psi_2,\dots,\psi_N)^T$ with
\begin{equation}
  \psi_j(n,m,l)=\sum^{2}_{s=0}\varrho^{(0)}_{j,s}
  (-\omega_s (k_j))^l (a-\omega_s(k_j))^n (b-\omega_s( k_j))^m,
\label{psi-gen-HB}
\end{equation}
where $\omega_0(k)\equiv k$, $\omega_i (k)$ for $i=1,2$
are def\/ined in~\eqref{omega-j} and $\varrho^{(0)}_{j,s}$
are constants.
Here a $N$th order Casoratian of the column vector
\begin{equation*}
\psi(n,m,l)=(\psi_1(n,m,l),\psi_2(n,m,l),\dots,\psi_{N}(n,m,l))^T
\end{equation*}
w.r.t.\ shift variable $l$, is def\/ined by
\begin{equation*}
  %\label{eq:C-gen}
  C_{n,m}(\psi;\{l_i\})=
  |\psi(n,m,l_1),\psi(n,m,l_2),\dots,\psi(n,m,l_N)|.
\end{equation*}
Usually we write this in the shorthand notation
\cite{Freeman-Nimmo-KP} $|l_1,l_2,\dots,l_N|$, and in particular for
con\-secutive sequences we use $\h{M}\equiv 0,1,\dots,M$.

The proof for \eqref{bil-HDB-d} is  similar to the one  in \cite{HZ10}.
In fact, the Casoratian column vector $\psi$ def\/ined in \eqref{psi-gen-HB}
satisf\/ies the shift relation
\begin{subequations}
\label{cas-cond-b2}
\begin{gather}
 \t\psi-\b\psi=a\psi,\qquad \h\psi-\b\psi=b\psi,\label{cas-cond-b2-a}\\
 \Gamma\psi=\b{ \b {\b \psi}}+b_0 \b {\b \psi},\label{cas-cond-b2-b}
\end{gather}
\end{subequations}
where $\Gamma$ is some $N\times N$ matrix.  We note that the relation~\eqref{cas-cond-b2-a} is the same as the one in~\cite{HZ10}, and it is
suf\/f\/icient for proving the f\/irst three equations in~\eqref{bil-HDB-I},
so here we do not consider them further.  The relation~\eqref{cas-cond-b2-b} is dif\/ferent from the one in~\cite{HZ10}.  Using
it one can generate an explicit form for the identity
$(\mathrm{Tr}(\Gamma)\dt{f})\dh{f}=(\mathrm{Tr}(\Gamma)\dh{f})\dt{f}$
(c.f., Appendix of~\cite{HZ10}),
\begin{gather*}
  a^{N-2}\dt{f}\Bigl[|\h{N-5},N-3,N-2,N-1,\dh \psi(N-2)|  - |\h{N-4},N-2,N,\dh \psi(N-2)|\\
  \qquad \quad {}+|\h{N-3},N+1,\dh \psi(N-2)|   + b^{N+1}\dh{f}+g-b  f +b_0 b^{N-2}( \dh{s}-\dh{h})\Bigr] \\
\qquad{} -b^{N-2}\dh{f}\Bigl[|\h{N-5},N-3,N-2,N-1,\dt \psi(N-2)|  - |\h{N-4},N-2,N,\dt \psi(N-2)|\\
\qquad \quad {} +|\h{N-3},N+1,\dt \psi(N-2)|   + a^{N+1}\dt{f}+g-a f +b_0 a^{N-2}(
 \dt{s}-\dt{h})\Bigr]=0,
 \end{gather*}
by which one can then verify \eqref{bil-HDB-d}.

\section{Conclusions}
In this paper we have discussed the various soliton solutions of the
recently discovered \cite{JH11} deformation \eqref{DB-cp} of the
lattice Boussinesq equation \eqref{DB}. It is well known that for
Boussinesq-type equations the soliton solution is formed by a combination of
three plane waves, but in the standard case with $b_0=0$ the parts
merge and produce a kink with oscillations. For $b_0\neq 0$ the three
components are more independent and the solutions look like resonating
two-kink solutions with three dif\/ferent asymptotic levels. Depending
on the choice of parameters there can be oscillations on the
transition region between levels.

\subsection*{Acknowledgments}
One of the authors (JH) was partially supported by Ville de Paris
within the ``Research in Paris'' program.  DJZ was supported by the
NSFC (No.~11071157).  This project is also partially supported by the
Shanghai Leading Academic Discipline Project (No.~J50101).

\pdfbookmark[1]{References}{ref}
\LastPageEnding

\end{document}